\documentclass[sigconf]{acmart}

\usepackage{array}
\usepackage{hyperref}
\usepackage[nolist]{acronym}
\usepackage{booktabs}
\usepackage{multirow}
\usepackage{subcaption}
\usepackage{enumitem}
\usepackage{wrapfig}
\usepackage{balance}
\usepackage{hyperref}

\AtBeginDocument{%
  }

\setcopyright{None}
\copyrightyear{2024}
\acmYear{2024}
\acmDOI{XXXXXXX.XXXXXXX}

\acmConference[AutomotiveUI '24 Adjunct]{16th International Conference on Automotive User Interfaces and Interactive Vehicular Applications}{September 22--25, 2024}{Stanford, CA, USA}

\acmISBN{978-1-4503-XXXX-X/18/06}

\copyrightyear{2025}
\acmYear{2025}
\setcopyright{rightsretained}
\acmConference[AutomotiveUI Adjunct '25]{17th International Conference on Automotive User Interfaces and Interactive Vehicular Applications}{September 21--25, 2025}{Brisbane, QLD, Australia}
\acmBooktitle{17th International Conference on Automotive User Interfaces and Interactive Vehicular Applications (AutomotiveUI Adjunct '25), September 21--25, 2025, Brisbane, QLD, Australia}
\acmDOI{10.1145/3744335.3758500}
\acmISBN{979-8-4007-2014-7/2025/09}

\begin{document}

\title[Quo-Vadis Multi-Agent Automotive Research?]{Quo-Vadis Multi-Agent Automotive Research? Insights from a Participatory Workshop and Questionnaire}

\author{Pavlo Bazilinskyy}
\email{p.bazilinskyy@tue.nl}
\orcid{0000-0001-9565-8240}
\affiliation{
  \institution{Eindhoven University of Technology}
  \city{Eindhoven}
  \country{The Netherlands}
}

\author{Francesco Walker}
\email{f.walker@fsw.leidenuniv.nl}
\orcid{0000-0002-7728-3542}
\affiliation{
  \institution{Leiden University}
  \city{Leiden}
  \country{The Netherlands}
}
\author{Debargha Dey}
\email{d.dey@tue.nl}
\orcid{0000-0001-9266-0126}
\affiliation{
  \institution{Eindhoven University of Technology}
  \city{Eindhoven}
  \country{The Netherlands}
}
\author{Tram Thi Minh Tran}
\email{tram.tran@sydney.edu.au}
\orcid{0000-0002-4958-2465}
\affiliation{
  \institution{The University of Sydney}
  \city{Sydney}
  \country{Australia}
}

\author{Hyungchai Park}
\orcid{0000-0002-7048-6363}
\email{hpark@stanford.edu}
\affiliation{
  \institution{Stanford University}
  \city{Incheon}
  \country{Republic of Korea}
}

\author{Hyochang Kim}
\orcid{0000-0002-4279-6620}
\email{hckim22@stanford.edu}
\affiliation{
  \institution{Stanford University}
  \city{Incheon}
  \country{Republic of Korea}
}

\author{Hyunmin Kang}
\orcid{0000-0003-2558-7744}
\email{neets@stanford.edu}
\affiliation{
  \institution{Stanford University}
  \city{Incheon}
  \country{Republic of Korea}
}

\author{Patrick Ebel}
\orcid{0000-0002-4437-2821}
\email{ebel@uni-leipzig.de}
\affiliation{
  \institution{ScaDS.AI, Leipzig University}
  \city{Leipzig}
  \country{Germany}
}

\renewcommand{\shortauthors}{Bazilinskyy et al.}

\begin{abstract}
The transition to mixed-traffic environments that involve automated vehicles, manually operated vehicles, and vulnerable road users presents new challenges for human-centered automotive research. Despite this, most studies in the domain focus on single-agent interactions. This paper reports on a participatory workshop (\textit{N} = 15) and a questionnaire (\textit{N} = 19) conducted during the AutomotiveUI '24 conference to explore the state of multi-agent automotive research. The participants discussed methodological challenges and opportunities in real-world settings, simulations, and computational modeling. Key findings reveal that while the value of multi-agent approaches is widely recognized, practical and technical barriers hinder their implementation. The study highlights the need for interdisciplinary methods, better tools, and simulation environments that support scalable, realistic, and ethically informed multi-agent research.
\end{abstract}

\ccsdesc[500]{Human-centered computing~Human computer interaction (HCI)}
\ccsdesc[500]{Computing methodologies~Modeling and simulation}

%
\keywords{Multi-Agent Research, Automotive, Interfaces, Automated Driving, Road Users}


\maketitle

\section{Introduction}
To operate safely, automated vehicles (AVs) must function as responsible social agents within complex mixed-traffic environments. These settings typically include a variety of agents including AVs, manually driven vehicles (MDVs), and vulnerable road users (VRUs). The goals and intentions of these agents often differ and can be conflicting. 
A key challenge in such environments is the variability in situation awareness (SA) between agents. SA refers to understanding what is happening in one's environment~\cite{Endsley1995}. When multiple road users interact, SA extends beyond individual cognition and becomes a shared construct. According to the Distributed Situation Awareness theory~\cite{stanton_distributed_2006}, traffic safety depends on shared and dynamically updated knowledge across all involved agents. Achieving this requires real-time multidirectional information exchange. Yet, much of the current research focuses on isolated user interactions, leaving the dynamics of multi-agent SA underexplored.

Real-world traffic involves complex social interactions such as group dynamics~\cite{jiang_joint_2018} and cooperative behaviors --- e.g., an AV warning a pedestrian about another approaching vehicle~\cite{ter_borg_future_2019}. These scenarios present challenges to the scalability and clarity of communication~\cite{dey2020taming, tran2023scoping}. For example, in situations with multiple pedestrians, an eHMI displaying ``Walk'' may be ambiguous with respect to its intended recipients~\cite{dey2020taming}.

Multi-agent settings further complicate joint decision-making under uncertainty, demanding intelligent systems capable of processing and interpreting inputs from multiple (human or non-human) agents. These systems must ensure coherence in behavior, such as avoiding contradictory signals. Computational models~\cite{lorenz2024computational} can contribute to this goal by capturing human perception, decision making~\cite{Ebel2023}, and cognitive processes~\cite{Oulasvirta2022}, allowing deeper insights into behavioral adaptations in dynamic traffic. Initial single-agent modeling efforts involving drivers and VRUs highlight the promise of such approaches and underscore the relevance of computational modeling in the advancement of multi-agent research~\cite{jokinen_predicting_2025, Wang2024}. 

Despite some progress, existing studies often rely on overly simplistic scenarios~\cite{kooijman_how_2019, fridman_walk_2019}, which do not capture the complexity of real-world multi-agent traffic scenarios. However, some efforts address this problem. \citet{bazilinskyy2020coupled} developed an open-source simulator that allows large-scale multi-user interaction in shared virtual traffic scenarios. This setup has been used to study collaborative behaviors such as pedestrian-AV eye contact~\cite{mok_stopping_2022} and eHMI-guided collision avoidance~\cite{bazilinskyy2022get}. Several recent works similarly showcased linked / distributed simulators that promote the investigation of multi-agent interaction research and enable immersive real-time interaction between users, offering rich insights into joint behavior~\cite{feng2023does, tran2024evaluating}.

From a cognitive psychology perspective, the influence of group dynamics on SA~\cite{9864310}, workload~\cite{lee2020workload}, and trust~\cite{walker2023trust} remains largely unexplored. \citet{moore_development_2010} took an initial step by analyzing group conversations during Tesla Autopilot rides, finding that most groups (94\%) experienced shared feelings of vulnerability and engaged in collaborative meaning-making to form shared mental models of AV behavior~\cite{momen_group_2023}. However, this work lacked objective behavioral measures.

We argue that a clear gap exists in the automotive interaction research literature concerning multi-agent traffic interaction, particularly from human factors and design perspectives. In this paper, we present initial findings from a participatory workshop that systematically explored the current state of multi-agent research and examined the key challenges and requirements for conducting rigorous studies in the context of automated mobility.

\section{Participatory Workshop}

\subsection{Method}
\begin{figure*}[h!]
\begin{center}
 \begin{subfigure}[t]{0.49\textwidth}
    \centering
    \includegraphics[width=\textwidth]{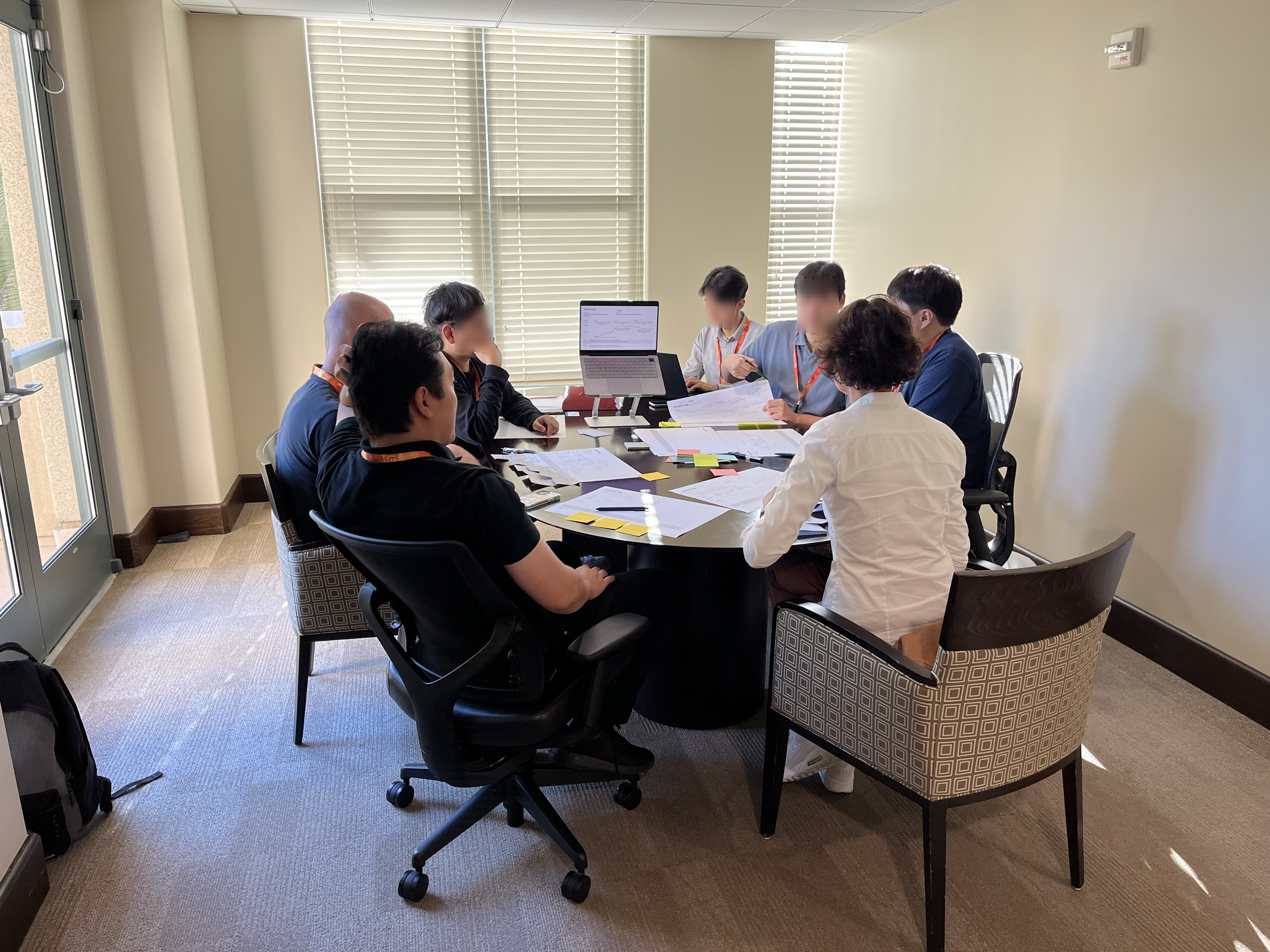}
    \label{fig1}
\end{subfigure}
 \begin{subfigure}[t]{0.49\textwidth}
    \centering
    \includegraphics[width=\textwidth]{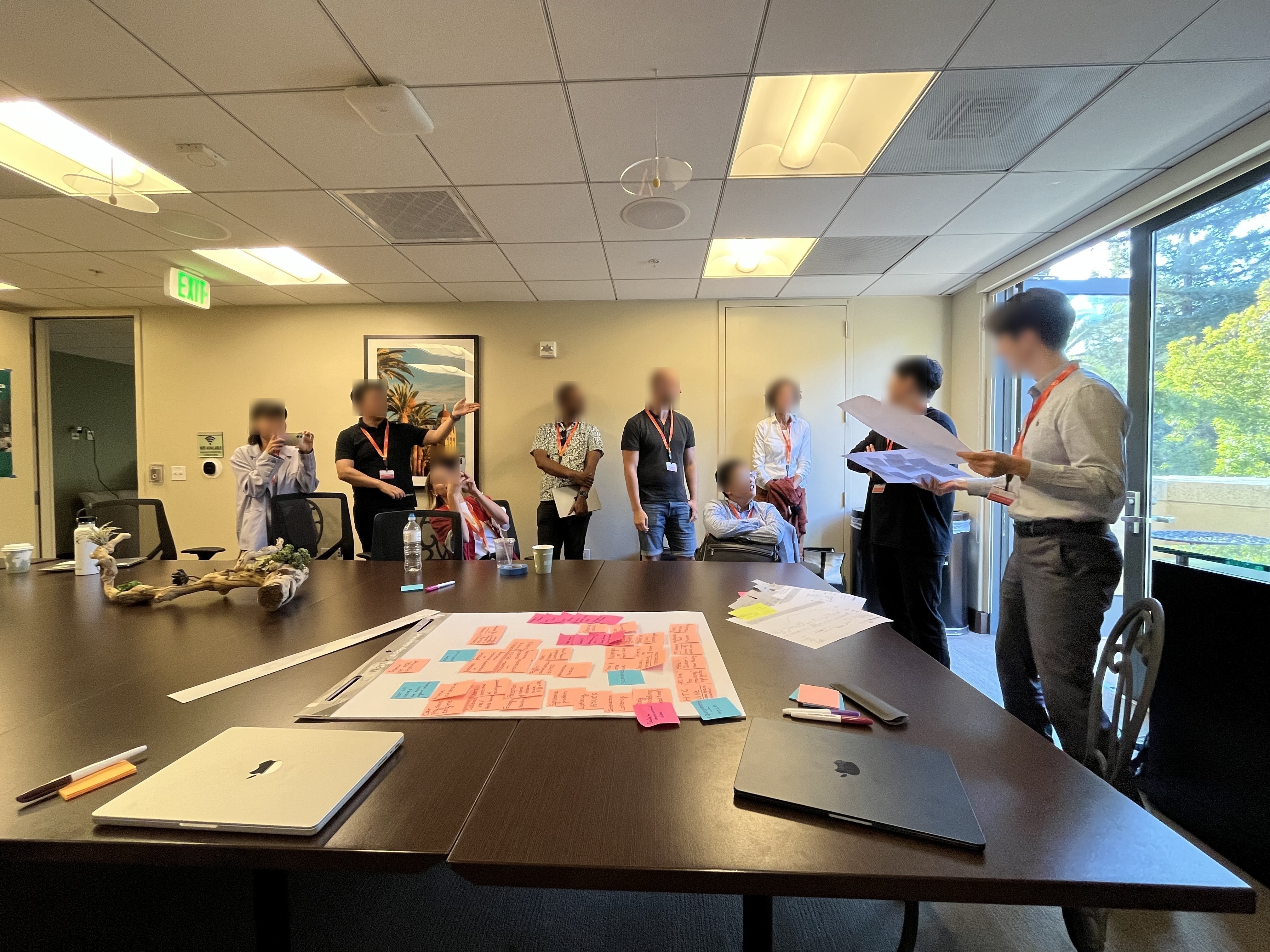}
    \label{fig2}
\end{subfigure}
\caption{Participants in the workshop engaged in one of the group work discussions (left) and the discussion at the end of the workshop (right). Faces are blurred for anonymity.} 
\label{fig:workshop}
\end{center}
\end{figure*}

The participatory workshop was organized as part of an academic conference on automotive user interfaces on 22 September 2024 during the AutomotiveUI conference in Stanford, CA, USA. Fifteen attendees of the conference participated in the workshop. The workshop (see Fig.~\ref{fig:workshop}) included an introduction, a presentation on existing multi-agent automotive research, and two group work sessions separated by a break on the topics of A. ``Multi-agent research in real-world settings'', B. ``Multi-agent research in VR simulations'' and C. ``Multi-agent modeling research in automotive'', live demos of the coupled simulator \cite{bazilinskyy2020coupled}. During group sessions, participants used sticky notes to summarize the main points. The workshop ended with group presentations and a summary. The study was approved by the Ethics Review Board of Eindhoven University of Technology and the participants (both in the participatory workshop and questionnaire) gave their informed consent to use their data for research purposes.

\subsection{Results}
\subsubsection{Multi-agent research in real-world settings}
This group explored both the potential and the challenges associated with conducting multi-agent research in naturalistic settings. First, through analyzing AV driving videos and structured scenario mapping exercises, participants identified various scenarios where multi-agent interactions could realistically occur (see Fig.~\ref{fig:workshop-scenrarios}). They highlighted that research on communication in mixed-traffic scenarios needs to move beyond hypothetical interactions and should address genuine human emotions and behaviors, such as hesitation, anxiety, and trust, which significantly influence interactions in real-world contexts.

The discussion highlighted the \textbf{importance of social and traffic norms in shaping human behavior in daily traffic}. For instance, pedestrians often decide to cross or hesitate based on the subtle movements of vehicles. Similarly, drivers engage in informal negotiations about the right-of-way at intersections. These examples suggest that human road users are highly dependent on shared, often tacit, understandings to navigate traffic interactions. Participants said that multi-agent research can help identify and systematize these norms and explore how eHMIs could be used to reinforce or even subtly reshape them. Drawing from real experiences, such as the ride of a participant in a Waymo robotaxi in San Francisco, CA, USA, that struggled to navigate an unexpected construction zone, they highlighted the unpredictability of real-world environments and the need for systems that can accommodate and respond to such complexity. Sticky notes (see Fig.~\ref{fig:workshop-notes}) highlighted themes such as dependency on the interaction context, communication between different types of agents, and the need to design systems that can handle and communicate uncertainty.

Particular attention was paid to the \textbf{limitations of conducting multi-agent research in a naturalistic setting}. The participants of the group discussion reached the consensus that multi-agent research in the real world is expensive, time-consuming, hard to set up, and that tools can be difficult to acquire. In addition, they acknowledged significant safety concerns and emphasized the complexity that arises from variations in national traffic regulations and differing cultural perceptions of traffic interactions. Participants concurred that a thorough understanding of legal and societal nuances is crucial but presents significant barriers to effectively executing multi-agent studies in naturalistic settings.

\subsubsection{Multi-agent research in VR simulations}
The second group explored the potential of virtual reality (VR) as a powerful yet sometimes limited tool for investigating multi-agent scenarios in automotive contexts. They began by discussing the types of multi-agent scenarios currently being simulated in VR, such as pedestrian-vehicle interactions or collaborative tasks in shared spaces. Participants found that many of the challenges they encountered were not specific to multi-agent setups, but inherent to VR itself. A central question was how to make VR-based interactions feel more realistic and natural without losing experimental control. The term 'pseudo-naturalistic interaction' was used to describe a desired balance: \textbf{simulations that feel authentic to users while remaining controlled and measurable.}. Achieving this balance can be particularly challenging in multi-agent scenarios, where the behavior of one agent directly influences others, resulting in a high degree of variability.

Another discussion revolved around the types of equipment used in such simulations. The group argued for \textbf{a flexible approach that includes both VR and non-VR input methods} (e.g., keyboard, mouse, touch interfaces) to improve accessibility and realism. It suggests a need for a modular multi-agent framework.
Yet another debate centered on whether participants in a multi-agent study should be physically or virtually aware of each other, for example, occupying the same physical/ simulation space, versus being isolated, and how this awareness could affect their behavior and the study results. The sticky notes of the participants (see Fig.~\ref{fig:workshop-notes}) emphasized the importance of technical flexibility, scalability, and social presence in future simulations. Participants also expressed interest in refining interaction models to incorporate more subtle social cues, improving motion realism, and investigating how environmental context influences agent behavior. The group concluded that while VR offers immense potential for studying complex human-vehicle interactions, its design must be carefully calibrated to produce valid and transferable insights.

\subsubsection{Multi-agent modeling in automotive interaction research}
The third group tackled the complex question of whether current modeling approaches in automotive research are sufficient for understanding and designing multi-human interactions in AVs. The discussion opened with reflections on the limitations of single-agent models, which often ignore group dynamics, and quickly moved to examples illustrating why more complex modeling is needed. For example, a group of pedestrians crossing a street may behave differently depending on whether one person or multiple people initiate(s) jaywalking. The system response must not only anticipate individual decisions but also account for the behavior of the emerging group. Participants argued that such interactions cannot be effectively captured using simple replication of one-agent models and called for more nuanced, multi-agent modeling approaches that can account for interdependent decision making.

A range of modeling strategies were discussed, including computational rationality (e.g., modeling how a person’s ability to predict future states impacts system responses) and rule-based attention management systems that help AVs infer where a person should be looking. The group also explored whether particle or fluid simulations might be sufficient to model crowd behaviors without resorting to full multi-agent architectures. This led to a debate where some felt that such methods could replicate general movement patterns, while others stressed that edge cases, such as a child darting into the street, require agent-level understanding. Use cases such as adjusting the HVAC (Heating, Ventilation, and Air Conditioning) settings to balance comfort for multiple passengers in a shared AV highlighted the growing need for models that reflect competing user needs and preferences.

Participants also acknowledged the tension between academic interest in theoretical models and the pragmatic concerns of industry, which may accept functional systems without fully understanding their internal dynamics. The session notes highlighted themes such as group dynamics, trade-offs between model complexity and tractability, and ethical considerations in shared decision making. The group ultimately concluded that while the current state of multi-agent modeling is still developing, it will be critical for supporting the safe, adaptive, and socially intelligent behavior required of future AVs.

\section{Questionnaire}

\subsection{Method}
To evaluate the experience with and attitudes toward multi-agent research, we conducted an online questionnaire. We also collected information to assess whether the participants are knowledgeable and have sufficient experience in multi-agent research. The complete questionnaire can be found in the supplementary material. The questions tackle demographics, experience with the conference, experience with multi-agent research, and questions about barriers to conducting multi-agent research and functionalities participants would expect from software and tools that aim to enable multi-agent research. 

Nineteen participants were recruited through opportunity sampling in two phases associated with the AutomotiveUI '24 conference. Recruitment occurred (1) before the workshop via a questionnaire completed by registered workshop participants, (2) during the event itself through QR code flyers and word-of-mouth efforts by workshop attendees, and (3) at a social gathering held during the conference.
The completion times of the questionnaire ranged from 3 minutes 23 seconds to 37 minutes 32 seconds (\textit{Median} = 8:04 min; \textit{Mean} = 10:12 min; \textit{SD} = 6:42 min).

Of the 19 participants, 17 are from academia and 2 from industry. Their roles span faculty (\textit{n} = 5), staff researchers (\textit{n} = 4), postdoctoral researchers (\textit{n} = 3), graduate students (\textit{n} = 4), PhD candidates (\textit{n} = 2), and one part-time research assistant. Although four participants are first-time attendees at the conference, most have attended previously, including eight who have participated three or more times, indicating strong engagement with the community.
Most of the participants conduct interdisciplinary research on human interaction with AVs, focusing on topics such as user experience, behavior, trust, interface design, communication, uncertainty, ethics, and psychology. The vast majority primarily rely on experimental methods such as simulations, virtual reality, and user studies, while only a few engage in interaction modeling and, even then, usually as a secondary or collaborative effort. Anonymized questionnaire data are available in \autoref{sec:supplemtary}.

\subsection{Results}
\begin{figure*}[htb]
\footnotesize
	\subfloat[How often have respondents worked with specific kinds of data in the past?\label{subfig:kindsData}]{%
		\includegraphics[width=0.47\linewidth]{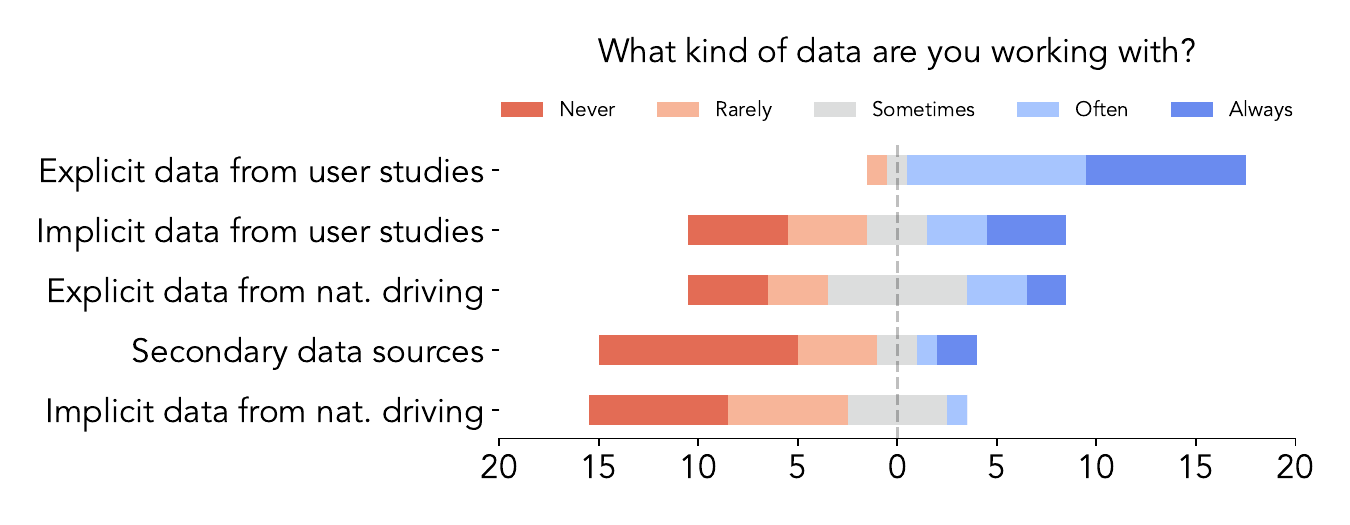}
	}
	\hfill
	\subfloat[How often have respondents worked with specific types of data in the past?\label{subfig:typesData}]{%
		\includegraphics[width=0.47\linewidth]{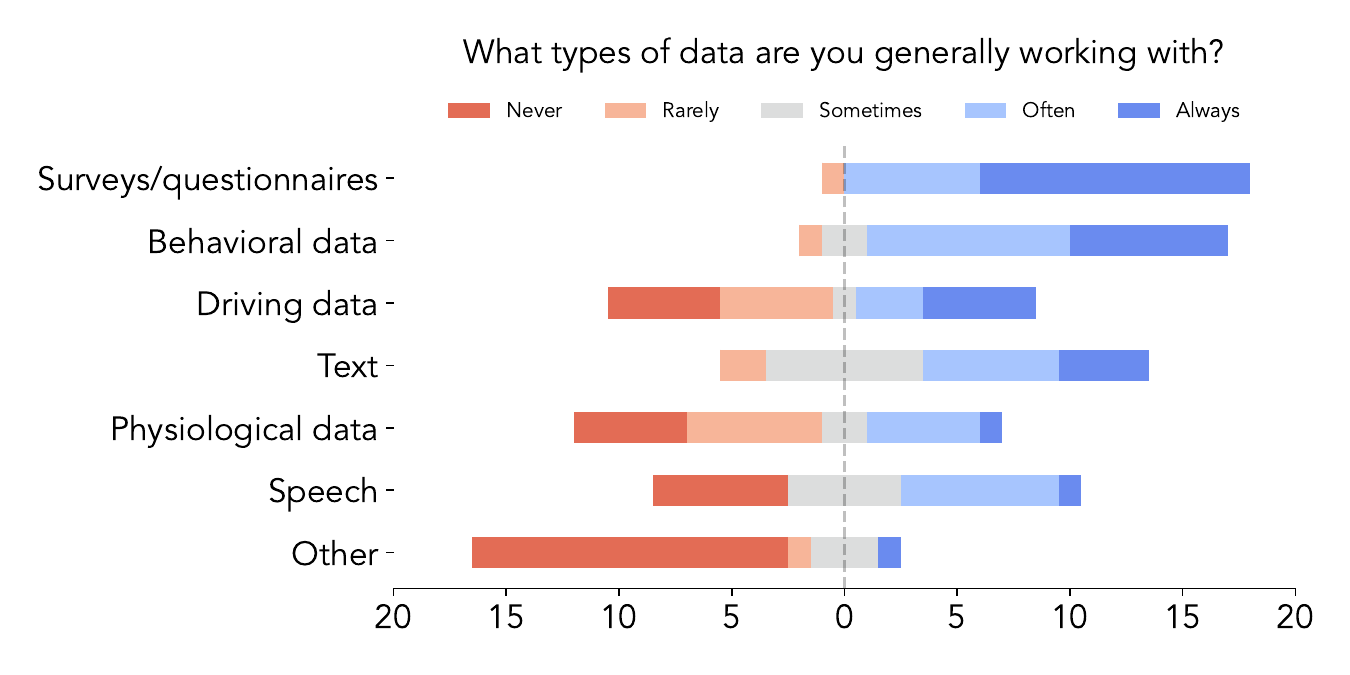}
	}
	\hfill
	\subfloat[Barriers for conducting multi-agent research.\label{subfig:barriers}]{%
		\includegraphics[width=0.47\linewidth]{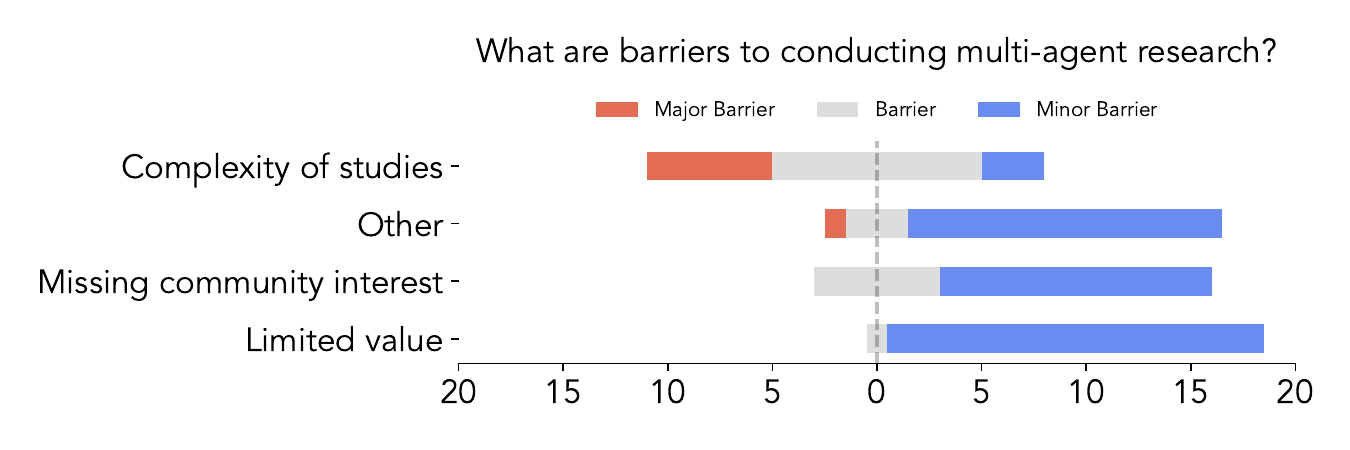}
	}
 	\hfill
	\subfloat[Features respondents expect from tools and software for multi-agent research.\label{subfig:services}]{%
		\includegraphics[width=0.47\linewidth]{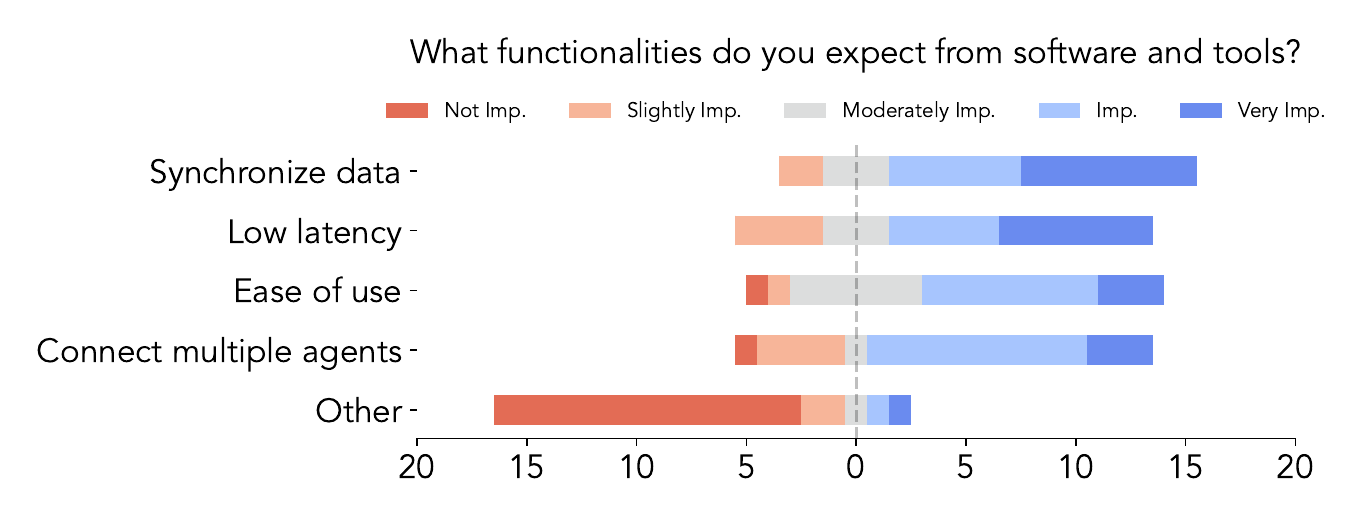}
	}
        \hfill
	\subfloat[Frequency of respondents conducting multi-agent research.\label{subfig:frequency}]{%
		\includegraphics[width=0.47\linewidth]{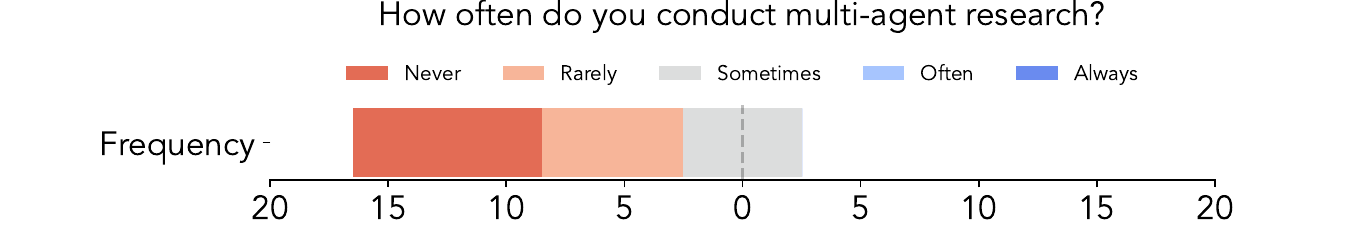}
	}	
\caption{Results from the questionnaire.}
\Description[Five Horizontal stacked bar charts]{Five horizontally stacked bar charts. Each bar chart visualizes either how often people work with specific types and kinds of data, what they consider the barriers, and how important they consider specific features of multi-agent research.}
\label{fig:surveyresults}
\end{figure*}

Most participants had experience in conducting user studies, working with both explicit and implicit data collected through controlled experiments (see \autoref{subfig:kindsData}).
However, few had experience with data from naturalistic driving studies (see \autoref{subfig:kindsData}), and none regularly conduct multi-agent research, with nearly half of the participants having never engaged in any multi-agent research (see \autoref{subfig:frequency}). Overall, the results show that participants only have limited experience conducting multi-agent research.

Respondents \textbf{recognize the importance of multi-agent research} and do not see a lack of community interest or personal appreciation as barriers to engaging in it.
They identified a wide range of potential use cases for multi-agent approaches in automotive and mobility research. Many (\textit{n} = 7) highlighted traffic scenarios in which the actions of one agent influence others, such as the dynamics of the group at intersections or traffic lights. P1 described the relevance of multi-agent research saying that \textit{``everything is almost always multi-user''}.
However, \textbf{the perceived complexity of multi-agent research presents a significant challenge}. Approximately 75\% of the respondents identify the complexity involved in designing, conducting and analyzing multi-agent studies as a barrier or a major barrier (see \autoref{subfig:barriers}).
When asked what they would need to conduct multi-agent research, respondents most frequently cited access to appropriate technology and hardware (\textit{n} = 8), followed by the need for methodological expertise and a strong theoretical foundation (\textit{n} = 4) as key enablers.
This indicates that methodological and technological difficulties are the main obstacles limiting the adoption of multi-agent approaches in automotive user research.

\section{Discussion}
The results of the questionnaire confirmed previous research~\cite{tran2023scoping} that multi-agent research in the automotive domain remains underexplored, with complexity identified as the main barrier. The focus group discussions revealed several promising directions that the community can explore in parallel, either to identify the most effective approaches or to generate complementary insights.

The first approach was grounded in real-world observations and involved capturing insights through scenario-based mapping~\cite{howard2014journey}. This method not only helps break down the complexity of traffic interactions but also illustrates how design research can contribute to multi-agent studies. By surfacing tacit norms such as hesitation, negotiation, and informal coordination, scenario mapping demonstrates the value of situated interpretive approaches. Related work has explored similar directions; for example, \citet{gao2024roleswitch} applied a role-switching method in which participants considered a traffic scenario from multiple perspectives (i.e., pedestrian, cyclist, driver of an MDV and passenger of an AV) to support understanding and design exploration. These examples suggest a broader potential for design research methods, such as speculative prototyping, co-design, and critical reflection, to contribute to the study and shaping of multi-agent dynamics.

The second and third approaches emphasized technical strategies. One focused on developing pseudo-naturalistic simulations and device-agnostic interaction setups to support more realistic and inclusive studies. This direction reflects growing research interest in moving beyond tightly controlled VR scenarios, given that restricted setups can influence pedestrian behavior and interaction patterns with AVs~\cite{tran2024advancing}. This shift towards more ecologically valid simulations raises a critical question: What constitutes a 'realistic' study. Although immersive virtual environments are often valued for their visual fidelity, realism is ultimately determined less by graphic detail and more by the degree to which users feel present within the virtual setting~\cite{slater2009place}. Crucially, factors such as the affective content of the experience and the fluidity of users’ movements and interactions within the environment play a greater role in enhancing presence than visual quality alone~\cite{banos2004immersion}. Thus, technical efforts to improve 'pseudo-naturalistic interactions' should prioritize the sense of presence over purely visual or immersive elements.

Alongside these simulation-based strategies, a complementary approach highlighted the potential of computational modeling to derive predictive insights from complex multi-agent scenarios. While these simulation and modeling strategies offer promising technical directions, it is worth noting that despite our efforts to cover a broad range of multi-user contexts, the discussion often focused on interactions between AVs and VRUs (e.g., pedestrians, cyclists). As a result, little attention was paid to a different but equally important context: interactions within the vehicle, among passengers sharing an automated ride. Although group dynamics has been recognized as a relevant factor in AV contexts, their role inside the vehicle remains notably underexplored.

Together, these approaches are primarily technical in nature and aim to improve systems, infrastructure, and methods. Data collected from simulation-based studies, such as VR experiments, may also serve as useful input for modeling efforts, particularly when grounded in scenarios that reflect real-world complexity. Within the AutoUI community, ongoing efforts toward open science~\cite{ebel2024changing} could support this direction by allowing greater data sharing and methodological transparency. In particular, computational modeling has already shown promise in evaluating UI in vehicles~\cite{lorenz2024computational}. This suggests core modeling techniques could potentially be adapted to evaluate external UIs within multi-agent traffic scenarios. As engineering- and implementation-oriented responses, these approaches contribute to the development of a technical foundation for future empirical and modeling work.

Furthermore, given that road environments involve multiple heterogeneous agents with varying goals, perceptual capacities, and levels of SA, leading to emergent dynamics like third-party interference, cascading hesitation, and negotiated right-of-way, researchers recognize that a shift toward multi-agent simulation is needed. This premise is grounded in Endsley’s SA model~\cite{Endsley1995}, which highlights the importance of perceiving, comprehending, and projecting within dynamic systems, and Hutchins’ theory of distributed cognition~\cite{hutchins_cognition_1995}, which emphasizes that awareness and decision-making are shaped by interactions across people and artifacts. In traffic, SA is not solely individual but often distributed and interdependent, making it critical to move beyond isolated dyads toward methods that reflect the complex, social, and adaptive nature of traffic behavior.

The discussion and insights presented in this paper are based on a single participatory workshop (\textit{N} = 15) and a small-scale questionnaire (\textit{N} = 19), with participants primarily from academic settings. The topics for discussion during the workshop were selected a priori by the organizers, drawing on their respective areas of expertise. The research directions outlined remain conceptual and untested in this study. Future work could build on these early findings through in-depth empirical evaluations, broader stakeholder engagement, and efforts to integrate diverse methodological approaches into practice.

\section{Supplementary Material}
\label{sec:supplemtary}
The anonymized results of the questionnaire and the analysis code are available at \url{https://doi.org/10.4121/40d7a8c5-2c68-4681-8a15-a224be5ca1fe}. The material is made public to support open science practices in the AutomotiveUI community \cite{ebel2024changing}.


\bibliographystyle{ACM-Reference-Format}
\bibliography{references,bazilinskyy}


\appendix
\renewcommand{\thefigure}{A\arabic{figure}}
\setcounter{figure}{0} 


\begin{figure*}[h!]
\begin{center}
 \begin{subfigure}[t]{0.49\textwidth}
    \centering
    \includegraphics[width=\textwidth, height=9cm]{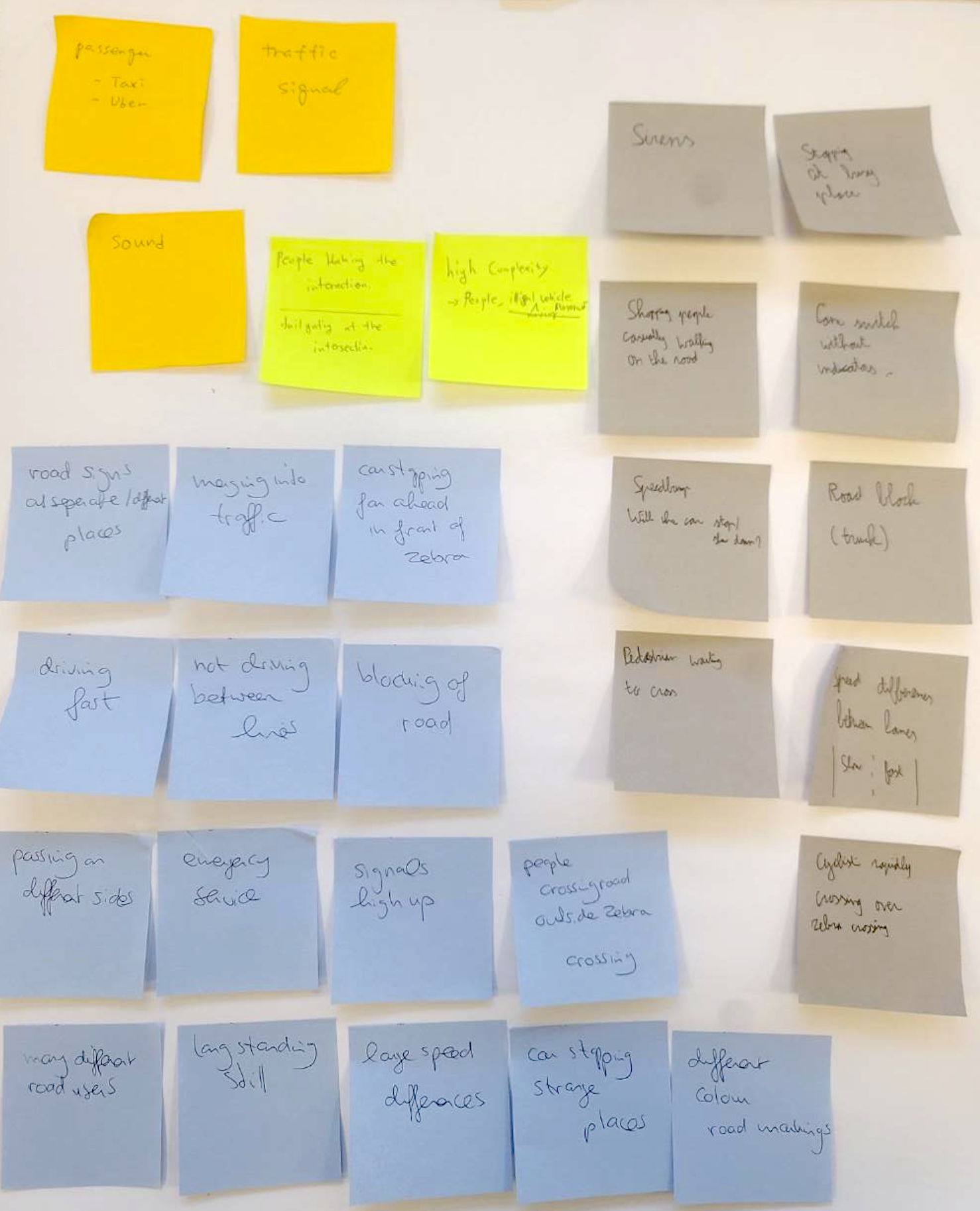}
    \label{fig1}
\end{subfigure}
 \begin{subfigure}[t]{0.49\textwidth}
    \centering
    \includegraphics[width=\textwidth, height=9cm]{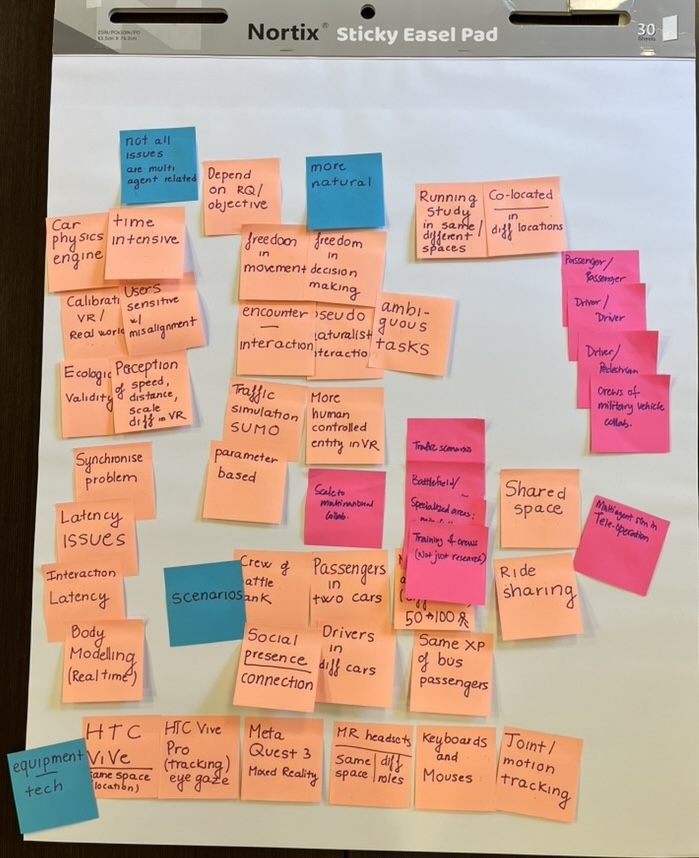}
    \label{fig2}
\end{subfigure}
\caption{Notes from discussion A. Multi-agent research in real-world settings (left) and B. Multi-agent research in VR simulations (right).} 
\label{fig:workshop-notes}
\end{center}
\end{figure*}

\begin{figure*}[!t]
    \centering
    \includegraphics[width=0.9\linewidth]{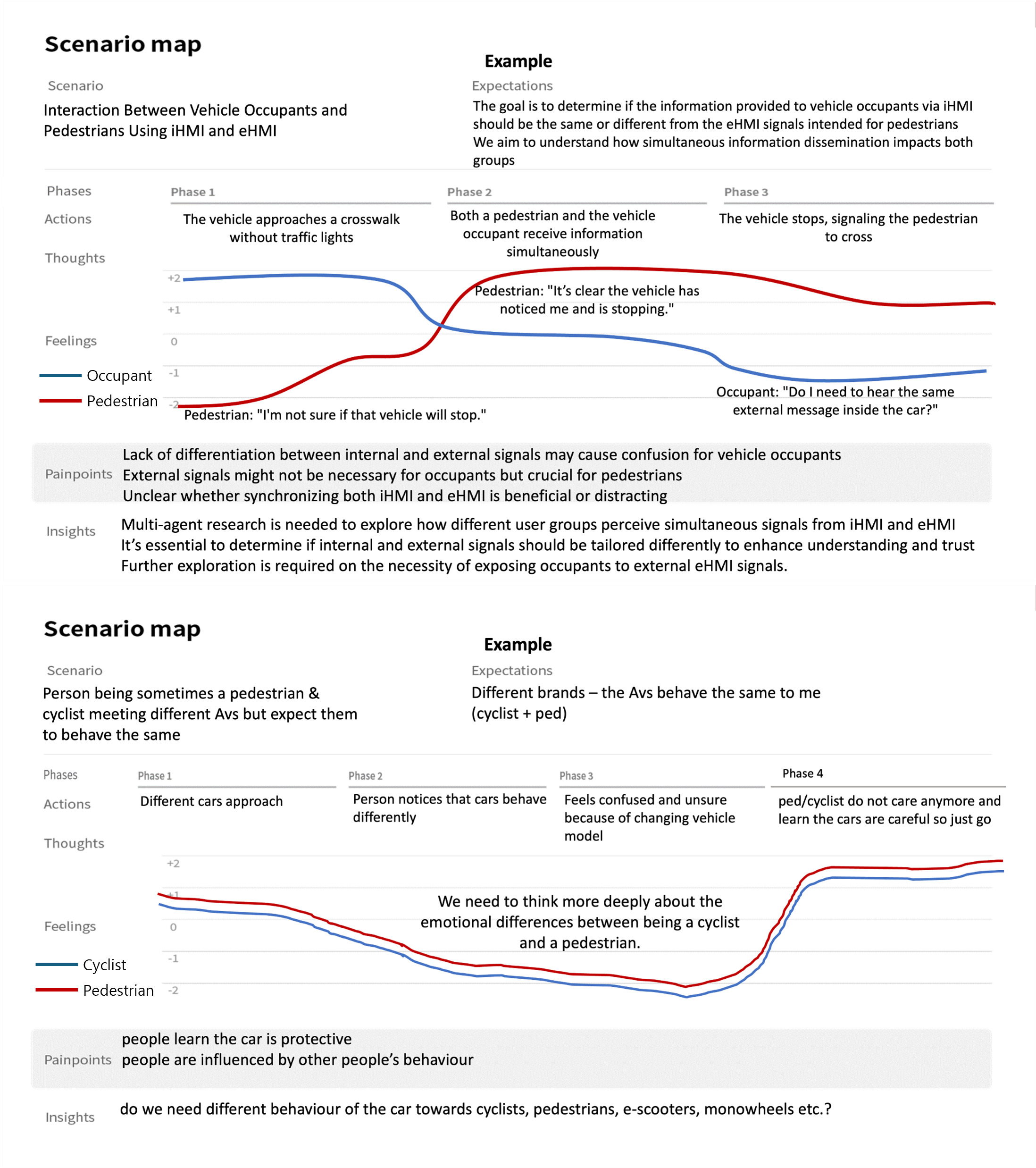}
    \caption{Two scenarios created by participants during discussion A. Multi-agent research in real-world settings. Participants divided the scenario into several phases and responded by drawing a continuous line to indicate one agent’s emotional state throughout the experience, with the Y-axis representing emotion scores ranging from -2 to +2. Thoughts were written in text, and differently colored lines represent the distinct emotional trajectories of two separate agents within the same scenario.}
    \label{fig:workshop-scenrarios}
\end{figure*}

\end{document}